\documentclass[prb,preprint]{revtex4-1} 

\usepackage{setspace}
\usepackage{amsmath}  
\usepackage{amsfonts} 
\usepackage{graphicx} 
\usepackage{braket}
\usepackage{caption}
\usepackage[dvipsnames]{xcolor}
\usepackage{romannum}
\usepackage{subfigure}
\usepackage{multirow}
\usepackage{bm}
\usepackage{float}
\begin{document}
\title{Numerical Simulation of Shell Model Single Particle Energy States using Matrix Numerov Method in Gnumeric Worksheet}
\author{Shikha Awasthi}
\author{Aditi Sharma}
\author{Swapna Gora}
\author{O.S.K.S Sastri}
 \email{Corresponding author: sastri.osks@hpcu.ac.in} % optional
\affiliation{Department of Physical and Astronomical Sciences, Central University of Himachal Pradesh, Dharamshala, Himachal Pradesh-176215, Bharat(India)}
\date{\today}

\begin{abstract}
Single particle energy states as described by nuclear shell model are obtained for doubly magic nuclei using Gnumeric worksheet environment. Numerov method rephrased in matrix form is utilised to solve time-independent Schr{\"o}dinger equation (TISE) within mean-field approximation, described by Woods-Saxon (WS) potential along with spin-orbit term, to obtain the single particle energies for both neutron and proton states. The WS model parameters are chosen from previous simulation results performed using matrix methods technique involving sine basis, where optimization was done w.r.t available experimental single particle energies for $^{208}_{82}Pb$ and $^{48}_{20}Ca$. In this paper, only the algorithm parameters, step size ‘h’ and matrix size ‘N’ are optimized to obtain the expected energy level sequence obtained using matrix methods. An attempt is made, by incorporating this tool within the framework of guided enquiry strategy (a constructivist approach to learning), to actively engage the students in assigning appropriate $J^\pi$ configurations for ground states of nuclei neighbouring the doubly magic ones. It has been observed that the ground state configurations could be better predicted when energy level sequences are known for all nuclei as compared to what is usually obtained from that of $^{208}_{82}Pb$ alone.
\end{abstract}
\maketitle
%%%%%%%%%%%%%%%%%%%%%%%%%%%%%%%%%%%%%%%%%%%%%%%%%%%%%%%%%%%%%%%%%%%%%
%% Start the main part of the manuscript here.
\setstretch{1.5}
%%%%%%%%%%%%%%%%%%%%%%%%%%%%%%%%%%%%%%%%%%%%%%%%%%%%%%%%%%%%%%%%%%%%%
\section{Introduction}
Shell model \cite{1} has been one of the most successful models to have explained evidence for magic numbers, that has emerged from binding energy data. While harmonic oscillator potential along with inclusion of spin-orbit term has been very effective in obtaining shell closures at magic numbers, actual energy level sequencing as seen from experimental data is better deduced by utilizing a rounded square well potential as demonstrated by WS potential \cite{2}. Even though square well and harmonic oscillator potentials are included in both under-graduate(UG) and post-graduate(PG) nuclear physics courses \cite{3}, they are not dealt beyond establishing the fact that magic numbers result due to $\bar{L}.\bar{S}$ splitting that gives rise to levels with higher j-values corresponding to a particular N-oscillator getting clubbed with those of a lower (N-1) oscillator. There is no way to judge the magnitude of splitting due to this spin-orbit coupling and hence different textbooks \cite{4,5,6} present varying energy level sequences which could lead to difficulties while assigning the ground state total angular momentum $J$ and spin-parities $(-1)^\ell$ for different nuclei. Another important lacuna in the pedagogy of topics in this subject is the lack of lab activities that enhance interaction with the content. Our physics education research (PER) group has been focussing on this much needed aspect and have developed various experimental \cite{7,8,9} and simulation \cite{10,11,12,13} activities to supplement the classroom lectures. With regard to single particle energy level structure, we have solved the TISE for WS potential along with spin-orbit term, utilizing matrix methods technique employing sine wave basis in Scilab \cite{10}. In spite of the simplicity of sine basis, the technique still requires determining integrals that appear in the matrix elements numerically. This makes it invariable to use a programming environment such as Scilab. So as to overcome this limitation and keep ease of simulating the problem using simple worksheet environment, matrix methods numerical technique has been replaced with Numerov method rephrased in matrix form \cite{14}.\\
In this paper, we utilise model parameters obtained through optimization, while solving TISE using matrix methods \cite{10}, with respect to available experimental single particle energies \cite{15}. The nuclear shell model with interaction potentials based on these model parameters, which are rephrased in appropriate choice of units, is described in Section II. A brief discussion on numerical Numerov method and its algorithm are given in Section III. The implementation details, for a typical example of $^{40}_{20}Ca$ using Gnumeric worksheet environment have been presented in a step by step approach in Appendix. The algorithm parameters are optimized to obtain convergence of single particle energies with those obtained using matrix methods \cite{10} and final results are discussed in Section IV along with our initial attempts at implementation of a pilot study, by incorporating this tool into guided enquiry strategy (GIS) framework. Finally, we draw our conclusions in Section V. 
\section{Nuclear Shell Model using Woods-Saxon potential:}
The modeling methodology \cite{16} has been described in great detail in our previous paper on Shell model simulation \cite{10}. So, only a brief description of the potentials rephrased in MeV and fm units are given here. In shell model, a nucleus of mass number A, consisting of N neutrons and Z protons has been modeled by the assumption that each nucleon experiences a mean field of central potential type due to rest of the nucleons. Woods-Saxon (WS) potential, which has typically a rounded square well shape, is one of the successful mathematical formulations given by
\begin{equation}
    V_{WS}(r)= \frac{V_0}{1 + \exp{\big(\frac{r - R}{a}\big)}}
    \label{eqn:1}
\end{equation}
where $V_0$ is the depth of the well, given by \cite{18}
\begin{align} 
V_{0}=\begin{cases}
-51+33((N-Z)/A) \quad MeV,
\quad \quad \quad \quad \quad  \textrm{for neutrons}\\
-51-33((N-Z)/A) \quad MeV ,
\quad \quad \quad \quad \quad  \textrm{for protons} 
\end{cases}
\label{eqn:2}
\end{align}
Here, R is the radius of the nucleus, empirically obtained as $R_0 A^{1/3}$, with value of $R_0$ being 1.28. \textit{a} is surface diffuseness parameter and is found to be 0.66 \cite{10}. \\
Next, interaction of spin of nucleon with orbital angular momentum of nucleon, as confirmed in experiments \cite{17}, has been modeled by spin-orbit potential, as
\begin{equation}
V_{ls}(r) = V_1 \Big(\frac{r_0}{\hbar}\Big)^2 \frac{1}{r} \frac{d}{dr} \Bigg[\frac{1}{1+\exp{\big(\frac{r-R}{a}\big)}}\Bigg] \textbf{(L.S)}
\label{eqn:3}
\end{equation}
Here, $\textbf{L.S} = [j(j+1) - \ell(\ell+1) - 3/4]\hbar^2$, where $\ell$ is orbital angular momentum quantum number, $j = \ell + s$ is total angular momentum quantum number and $s$ is spin angular momentum quantum number given by $1/2$ for nucleons. The model parameters are $V_1 = -0.44V_0$ \cite{18} and $r_0= 0.90$, a proportionality constant optimised \cite{10} to obtain the right energy level sequence.\\
In case of protons, Coulomb potential also needs to be considered and is given by
\begin{equation}
        V_c(r)= 
\begin{cases}
   \frac{(Z-1)e^2}{4 \pi \epsilon_0 r},\quad \quad
   \quad \quad \quad \quad\hspace{0.9cm} \textrm{for}\hspace{0.2cm} {r \ge R_c} \\
    \frac{(Z-1)e^2}{4 \pi \epsilon_0 R_c}\Big[\frac{3}{2}-\frac{r^2}{2R_c^2}\Big],
    \quad \quad \quad \quad \textrm{for}\hspace{0.2cm} {r \le R_c}
\end{cases}
\label{eqn:4}
\end{equation}
This potential is to be rephrased in MeV units. SO, it is multiplied and divided by electron rest mass energy, \cite{19} $m_{e}c^{2} = 0.511$ MeV 
to obtain
\begin{equation}
V_{c}(r)=\begin{cases}
          \frac{(Z-1)\ast 2.839\ast 0.511}{r}, \quad \hspace{1.8cm}\textrm{for}\  r \geq R_{c}\\
          \frac{(Z-1)\ast 2.839\ast 0.511}{R_{c}}\bigg[ \frac{3}{2}-\frac{r^{2}}{2R_{c}^{2}}\bigg], \quad \textrm{for}\  r \leq R_{c}
         \end{cases}
         \label{eqn:5}
\end{equation}
The radial TISE for central potentials is given by
\begin{equation}
\frac{d^{2}u(r)}{dr^{2}} + \frac{2\mu}{\hbar^{2}}\left(V(r) + \frac{\ell(\ell+1)\hbar^{2}}{2\mu r^{2}}\right)u(r) = E u(r)
\label{eqn:6}
\end{equation}
where V(r) is net interaction potential experienced by a neutron or a proton and second term inside bracket, resulting from solution of $\theta$-equation, is called as centrifugal potential, $V_{cf}(r)$. This is rephrased in MeV units, by multiplying and dividing it by $c^2$, so that 
\begin{equation}
  V_{cf}(r) =  \frac{\ell(\ell+1)\hbar^2c^2}{2\mu c^2 r^{2}}
  \label{eqn:7}
\end{equation}
The value of $\hbar c = 197.329$ MeV-fm and the reduced mass $\mu$ is given by:
\begin{align}
\mu = \begin{cases}
\frac{m_n*(Z*m_p+(N-1)*m_n)}{(Z*m_p+N*m_n)},\quad \textrm{for neutron}\\
\frac{m_p*((Z-1)*m_p+N*m_n)}{(Z*m_p+N*m_n)},
\quad \textrm{for proton}
\end{cases}
\label{eqn:8}
\end{align}
Here, $m_{p}=938.272$ and $m_{n}=939.565$ are masses of proton and neutron respectively, in units of MeV/c$^2$.
%%%%%%%%%%%%%%%%%%%%%%%%%%%%%%%%%%%%%%%%%%%%%%%%%%%%%%%%%%%%%%%%%%%%%
\section{Numerical Solution}
\subsection{Numerov technique in matrix form:}
Consider TISE for a general potential V(r), given by
\begin{equation}
 \frac{d^{2}u(r)}{dr^{2}}+k^{2}(r)u(r)=0
 \label{eqn:9}
\end{equation}
where
\begin{equation}
k^{2}(r)=\frac{2\mu}{\hbar^{2}}[E-V(r)-V_{cf}(r)]
\label{eqn:10}
\end{equation}
The advantage of this Eq. (\ref{eqn:9}) is that it is linear in 'u' having no first order derivative involved and is hence ideally suited for solving using Numerov method. The wave-function u(r) is expanded in Taylor series by explicitly retaining terms up-to $O(h^4)$ and is obtained to an accuracy of $O(h^6)$, \citep{14} 
\begin{equation}
 u(r+h) = \frac{2(1-\frac{5}{12}h^{2}[k(r)]^2)u(r)-(1 + \frac{1}{12}h^{2}[k(r-h)]^2)u(r-h)}{1 + \frac{1}{12}h^{2}[k(r+h)]^2} + O(h^{6})
 \label{eq:19}
\end{equation}
Discretizing x in steps of h as:\\ ${r_1, r_2, \ldots, r_{n-1}, r_n, r_{n+1}, \ldots, r_N}$.\\ 
Here, $r_{n} = r_{1} + n*h$ \\
Now expressing $u(r_n + h)$ as $u_{n+1}$, so on and similarly, $k(r_{n})$ as $k_{n}$, Eq. (\ref{eq:19}) can be written as
\begin{equation}
u_{n+1} = \frac{2(1-\frac{5}{12}h^{2}k_{n}^{2})u_{n} - (1 + \frac{1}{12}h^{2}k_{n-1}^{2})u_{n-1}}{1 + \frac{1}{12}h^{2}k_{n+1}^{2}} + O(h^{6})
\label{eq:20}
\end{equation}
Substituting from Eq. (\ref{eqn:10}),  $k_n^2 = \frac{2\mu}{\hbar^2}(E - V_n)$ into the above, clubbing the terms containing $V_n$ and $E$, it can be recast into the following form:
\begin{equation}\nonumber
 -\frac{\hbar^{2}}{2\mu}\frac{(u_{n-1}-2u_{n} + u_{n+1})}{h^{2}} + \frac{(V_{n-1}u_{n-1}+10V_{n}u_{n} + V_{n+1}u_{n+1})}{12} = E\frac{(u_{n-1}+10u_{n}+u_{n+1})}{12}
 \label{eq:21}
\end{equation}
One has to keep in mind that whatever may be the potential, if wave-function were to be normalised, it should tend to 0 as x tends to $\pm\infty$. This implies, one has to choose the region of interest (RoI) large enough to ensure that the wave-function dies down to zero in either direction. That is, x values are limited to an interval such as $[L_1,L_2]$, such that $u$ goes to 0 at both ends of the interval. More specifically, 
\begin{equation}
u_1 = u(r_1 = L_1) = 0  
\label{eq:22}
\end{equation}
 and 
 \begin{equation}
  u(r_N = L_2) = 0 
  \label{eq:23}
 \end{equation}
Expanding the above equation for all intermediate points ($j = 2,3,\ldots,N-1$), as in case of CDD formulation of TISE, one will get a matrix equation as
\begin{equation}
\left(\frac{-\hbar^{2}}{2\mu} A + BV \right)u = E Bu
 \label{eq:24}
\end{equation} 
where in, $u$ is a column vector 
$(u_2, \ldots, u_{n-1}, u_{n}, u_{n+1},\ldots, u_{N-1})$,\\ Similarly, V is a column vector
$(V_2, \ldots, V_{n-1}, V_{n}, V_{n+1},\ldots, V_{N-1})$, but is converted into a diagonal matrix with these values along its central diagonal. Matrices A and B are given by
\begin{equation}
A = \frac{I_{-1} - 2I_{0} + I_{1}}{h^{2}} 
 \label{eq:25}
\end{equation}
and
\begin{equation}
B=\frac{I_{-1} + 10 I_{0} + I_{1}}{12}  
 \label{eq:26}
\end{equation}
where $I_{p}$ is the matrix of $1$'s along $pth$ diagonal and zeros elsewhere. Both A and B are tridiagonal matrices.
Multiplying Eq. (\ref{eq:24}) by B$^{-1}$ on both sides, gives TISE as a matrix equation, utilising Numerov method,
\begin{equation}
\boxed{\left(\frac{-\hbar^{2}c^2}{2\mu c^2}B^{-1}A + V\right)u = Eu}
 \label{eq:27}
\end{equation}
with an error of $O(h^{6})$.\\
Notice that a factor of $c^2$ is introduced in both numerator and denominator to ensure the units are in MeV and fm as required in nuclear physics.\\
This is the final equation that needs to be solved numerically to get energy eigen-values and eigen-functions of a particle interacting with a given potential V(r). The imposition of boundary conditions as $u_{1} = u_{N} = 0$ is equivalent to embedding the potential of interest, inside an infinite square well potential. Finally, $ (N-2) \times (N-2) $ sub-matrices of A and B are utilised to solve for the energy eigen-values and their corresponding eigen-vectors. 
%%%%%%%%%%%%%%%%%%%%%%%%%%%%%%%%%%%%%%%%%%%%%%%%%%%%%%%%%%%%%%%%%%%%%%
\subsection{Algorithm for implementation in Gnumeric:}
A step by step approach to determining the single particle energies for protons of $^{40}_{20}Ca$ is presented.
\begin{enumerate}
\item \textbf{Initialisation of parameters:}
There are two sets of parameters:\\ 
(i) \textit{Physical system parameters}:\\ Object and interaction variables constituting as inputs and state variables which need to be determined, are outputs.\\
Figure (\ref{Figure 1}) shows object variables, interaction variables, algorithm variables, input variables and other variables required for the calculations.\\
\begin{figure*}[h!]
    \centering
    \includegraphics[width=16 cm, height=5.2 cm]{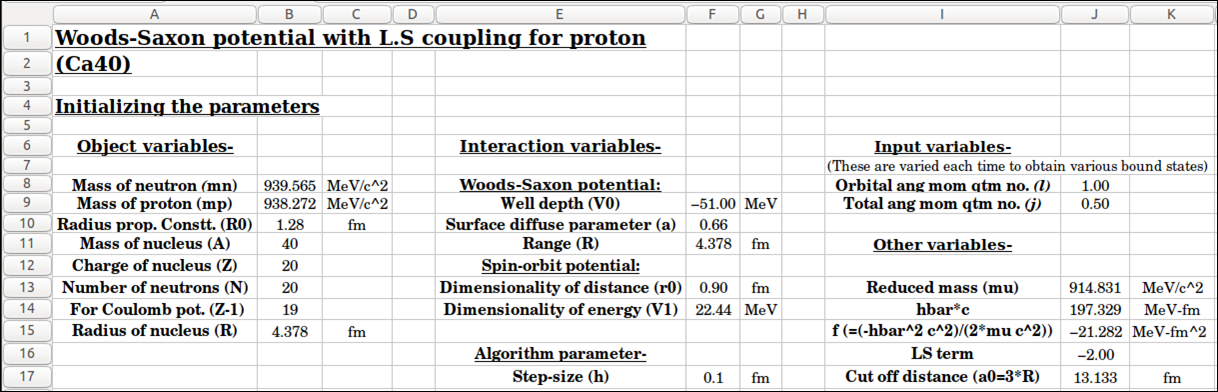}
    \caption{Initializing the parameters for the system}
     \label{Figure 1}
\end{figure*}
(ii) \textit{Algorithm parameters} that arise from discretization of continuous variables and limiting the infinitely large quantities to finite values such as region of interest. The step size is chosen as $h = 0.1$.
\item \textbf{Potential Definition:}\\
First values of 'r' are generated from $0.1$ to ($3R$) with step-size $h=0.1$ and total interaction potential $V(r)=V_{cf}(r)+V_{WS}(r)+V_{ls}(r)+V_c(r)$ for proton is obtained in Gnumeric worksheet. Figure \ref{Figure2} shows the plot of Woods-Saxon potential of $1s_{1/2}$, $1p_{1/2}$ and $1p_{3/2}$ states for proton, showing that inclusion of angular momentum on L.S coupling affects the Woods-Saxon potential. While centrifugal term pushes $\ell = 1$ levels above those of $\ell =0$, of $1s_{1/2}$ term, as it reduces the depth of potential, the spin-orbit term tends to act on width of potential. One can observe in Figure \ref{Figure2} that it decreases width of potential corresponding to $1p_{1/2}$ and increases it for $1p_{3/2}$. This would lead to raising the energy level corresponding to former to move upward while the later goes downward thus resulting in spin-orbit energy splitting.
%%%%%%%%%%%%%%%%%%%%%%%%%%%%%%%%%%%
%%%%%%%%%%%%%%%%%%%%%%%%%%%%%%%%%%%%%%%
\begin{figure*}[h]
\centering
\includegraphics[height=6cm, width=8cm]
{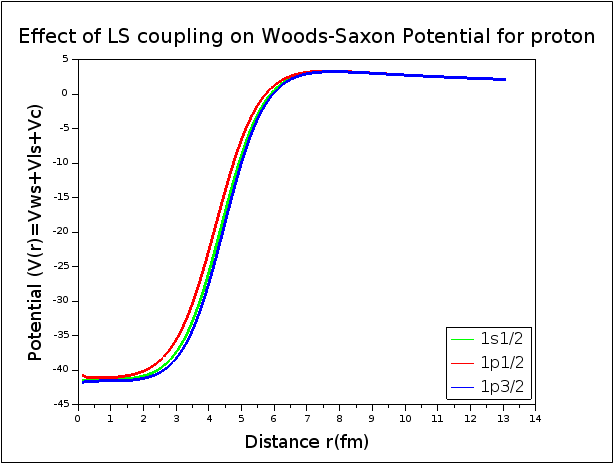}
\caption{Plots of $1s_{1/2}$, $1p_{1/2}$ and $1p_{3/2}$ states for proton, showing the effect of Spin-orbit interaction in Woods-Saxon potential.}
\label{Figure2}
\end{figure*}
%%%%%%%%%%%%%%%%%%%%%%%%%%%%%%%%%%%%%%%
\item \textbf{Determination of single particle energies:}\\
This step consists of generating the following matrices in successive sheets in Gnumeric. The size of the matrix will depend on :\\
(\textit{a}) B-matrix is generated in Sheet 1 with values 10/12 along its main diagonal and 1/12 along the first off diagonals on either side using an if statement. This is constant for all problems\\
(\textit{b}) The inverse of B, is obtained using minverse() command in sheet 2\\
(\textit{c}) A-matrix is generated similar to B-matrix
in sheet 3, by using if statement with appropriate formulae for factors f an g along the main and off diagonals respectively. 
(\textit{d}) $B^{-1}*A$ is generated in sheet 4 \\
(\textit{e}) V-matrix is generated in sheet 5, such that only diagonal elements get populated with those of net potential values \\
(\textit{f}) Generating Eigen values and eigen vectors:\\ The main feature of Gnumeric which makes it different from other worksheet environments such as MS-Excel or open Office Calc. is that, it has an eigen-value solver. Now, the eigen-values and corresponding eigen-vectors are obtained using eigen() command in sheet 6.
\item \textbf{Generating the energies for different $\ell$ and $j$ values:} Now, the bound state energies, those for which eigen values are negative, are tabulated for different values of $\ell$ and $j$.
The simulation is run for $\ell = 0$, s-states, in which case $j = 0.5$ alone exists. Then for $\ell = 1$, p-states, there are two values of j: 0.5 and 1.5. Similarly for $\ell = 2$,  d-states, $j$ takes values $1.5$ and $2.5$ and so on. This is continued till no bound states are obtained for particular $\ell$ and $j$ values.
\end{enumerate}

%%%%%%%%%%%%%%%%%%%%%%%%%%%%%%%%%%%%%%%%%%%%%%%%%%%%%%%%%%%%%%%%%%%
\section{Results and Discussions:}
In this section, we will first validate our approach to obtain energy eigen-values and eigen-functions using Woods-Saxon potential with $\bar{L}.\bar{S}$ coupling for both neutron and proton for doubly magic nucleus $^{40}_{20}Ca$. This is done by comparing our results with available experimental results and those obtained numerically by using matrix method with Fourier basis \cite{10} for different angular momentum values $\ell$ and j. The results are tabulated in Table  \ref{Table 1}.
%%%%%%%%%%%%%%%%%%%%%%%%%%%%%%%%%%%
\begin{table}[H]
    \begin{ruledtabular}
    \caption{Single particle shell model energy values for \textbf{Neutron states} and \textbf{Proton states} of doubly magic nucleus $^{40}_{20}Ca$ obtained by current work (\textit{using Matrix Numerov method} \cite{14}) with available experimental values \cite{15} and numerical values previously obtained by our group \cite{10}, (\textit{using Matrix method with Fourier basis}) for the highest occupied levels.}
    \label{Table 1}
    \begin{tabular}{llllllll}
{States} &\multicolumn{3}{c}{Proton states (MeV)}& {States} &\multicolumn{3}{c}{Neutron states (MeV)}\\
&{Exp.}&\multicolumn{2}{c}{Numerical values}&&{Exp.}&\multicolumn{2}{c}{Numerical values}\\
&{Ref\cite{15}}&{Ref\cite{10}}&{Current}&&{Ref\cite{15}}&{Ref\cite{10}}&{Current}\\
\hline
$1s1/2$&$\ldots$&$-30.49$&$-30.49$&$1s1/2$&$\ldots$&$-38.90$&$-38.90$\\
$1p3/2$&$\ldots$&$-21.68$&$-21.68$&$1p3/2$&$\ldots$&$-29.55$&$-29.55$\\
$1p1/2$&$\ldots$&$-19.04$&$-19.04$&$1p1/2$&$\ldots$&$-26.99$&$-26.99$\\
$1d5/2$&$-15.07$&$-12.19$&$-12.19$&$1d5/2$&$-22.39$&$-19.54$&$-19.54$\\
$2s1/2$&$-10.92$&$-8.14$&$-8.14$&$2s1/2$&$-18.19$&$-15.54$&$-15.54$\\
$1d3/2$&$-8.33$&$-6.85$&$-6.85$&$1d3/2$&$-15.64$&$-14.28$&$-14.28$\\
$1f7/2$&$-1.09$&$-2.33$&$-2.33$&$1f7/2$&$-8.36$&$-9.15$&$-9.15$\\
$2p3/2$&$0.69$&$1.00$&$1.00$&$2p3/2$&$-5.84$&$-5.42$&$-5.42$\\
$2p1/2$&$2.38$&$2.94$&$2.94$&$2p1/2$&$-4.20$&$-3.10$&$-3.10$\\
$1f5/2$&$4.96$&$5.37$&$5.37$&$1f5/2$&$-1.56$&$-1.19$&$-1.20$\\
\end{tabular}
    \end{ruledtabular}
\end{table}
%%%%%%%%%%%%%%%%%%%%%%%%%%%%%%%%%%%
%%%%%%%%%%%%%%%%%%%%%%%%%%%%%%%%%%%
It is observed that the obtained energy eigen values are in good agreement with experimental energy values of different neutron and proton single particle states for step size of h=0.1 thus validating our approach.
The distance parameter 'r' is discretized as per step size 'h' and its values are varied from $r = 0.1$ to $r = 3R$, where $R = R_{0}A^{(1/3)}$. Therefore, it is necessary to vary the size of matrix `N' for different A values for a chosen step size `h'. 
This simulation activity can be utilised as a tool to apply GIS\cite{20} of constructivist approach to learning and has been done as follows:
\subsection{GIS Implementation:}
The students have been taken through following six step process of GIS. All steps were implemented on online Moodle platform \cite{21} due to COVID-19 lockdown of university.
\begin{itemize}
\item \textbf{Initiation:} The matrix Numerov technique was already introduced before, for solving the harmonic oscillator potential. Next, it has been applied to solve for single particle energies of both neutrons and protons, as dealt with in this paper, for Woods-Saxon potential with spin-orbit potential. This has been explained and also demonstrated in two successive lab sessions.
\item \textbf{Selection:} The students were made to explore the binding energy and separation energies curves that they have plotted in previous sessions, to identify various double magic nuclei suitable for study. The doubly magic nuclei $^{16}_{8}O$, $^{48}_{20}Ca$, $^{56}_{28}Ni$, $^{100}_{50}Sn$, $^{132}_{50}Sn$  and $^{208}_{82}Pb$ were selected. The determination of single particle proton and neutron energy level sequences for each of these six double magic nuclei, are assigned to students by dividing them into $12$ groups. Each group is expected to obtain the energy level sequence of either neutrons or protons for the assigned nuclei by carefully following the simulation procedure.
\item \textbf{Presentation:} The students in each group could be asked to present their findings to the rest of the class so that everyone gets to know each other's experience.
Even though all the groups could get to expected level sequence by following the steps correctly, some of the students have not ensured reduction of step-size systematically. Hence, they have come up with lower accuracy for energies even though they obtained correct level sequence. They have been guided accordingly.
\item \textbf{Exploration:} The collective findings were shared with entire class and they were asked to explore nuclei adjacent to the magic numbers with one neutron or proton more and find out ground state $J^\pi$ configuration for each of them.  
\item \textbf{Formulation:} They were expected to formulate their findings based on right choice of level sequence and also figure out what they would get if they followed the level sequence given in their prescribed textbook \cite{4}.
\item \textbf{Collection:} To understand variation of energy level structure with mass number, students were asked to plot the compiled energy data for protons and neutrons as a function of mass number A. They were able to obtain plots similar to what have been presented in \cite{10}.
\item \textbf{Assessment:} They have assessed their formulated ground state angular momentum and spin configurations based on the experimental findings and thus validated their outcomes. They could be further asked to obtain the ground state configurations of nuclei slightly away from magic nuclei till the results obtained are in variance with those from experiments as an excercise.
\end{itemize}
\subsection{Outcomes from GIS:}
The results matching with experimentally available values \cite{15} were obtained for step-size '$h = 0.1$' for doubly magic nuclei $^{16}_{8}O$, $^{40}_{20}Ca$, $^{48}_{20}Ca$, $^{56}_{28}Ni$, $^{100}_{50}Sn$, $^{132}_{50}Sn$ and $^{208}_{82}Pb$. The value of N (to solve N $\times$ N matrix) for each of these nuclei are obtained as $96, 131, 138$, $145$, $177$, $194$ and $229$ respectively.
The obtained energy level sequences for neutrons and protons of these nuclei are given separately in tabular form in Tables \ref{Table 2} and \ref{Table 3}. The first four levels, $1s_{1/2},1p_{3/2}, 1p_{1/2}$ ad $1d_{5/2}$ are not shown, as there are no discrepancies found in the ordering of these levels across the periodic table due to our simulation. The numerically obtained energy values are found to match to two decimal places with those obtained using the matrix method approach \cite{10}. The $\chi^2$-value defined as relative mean-squared error\\
\begin{equation}
\chi^2 = \frac{1}{N}\sum_{i=1}^N \frac{(E_i^{expt}-E_i^{sim})^2}{|E_i^{expt}|}
\end{equation}
These are determined w.r.t experimental \cite{15} and are shown in Tables. In Tables \ref{Table 2} and \ref{Table 3}, energy level sequence obtained for lighter nuclei $^{16}O$ to $^{56}Ni$ are shown in first column and that for $^{100}Sn$ to $^{208}_{82}Pb$ in last column. It can be observed that the energy level sequence for lighter nuclei is different than that for heavier ones. The observed discrepancies in energy sequence are highlighted in red colour. \\
The discrepancy in level sequence for lighter nuclei w.r.t heavy nuclei is highlighted in blue colour in both tables; i.e for neutron and proton states. Further, it is found that in most of the textbooks at UG and PG level \cite{4,5,6,18,22,23}, energy level sequence given is different. Also, there is only single energy level sequence given for both neutrons and protons, and that too common for all mass ranges. But according to our calculations as well as Bohr and Mottelson book, there should be different energy level sequences for nuclei across the periodic table. \\
%%%%%%%%%%%%%%%%%%%%%%%%%%%%%%%%%%
%%%%%%%%%%%%%%%%%%%%%%%%%%%%%%%%%%
\begin{table}[H]
\begin{ruledtabular}
\caption{ Single particle shell model energy states for \textbf{Neutron} of doubly magic nuclei $^{16}_{8}O$, $^{48}_{20}Ca$, $^{56}_{28}Ni$, $^{100}_{50}Sn$, $^{132}_{50}Sn$ and $^{208}_{82}Pb$ obtained by using Matrix Numerov method}
\label{Table 2}
\begin{tabular}{llllllll}
{States}&\multicolumn{6}{c}{Numerical energy values(MeV)}&{States}\\
&{$^{16}_{8}O$}&{$^{48}_{20}Ca$}&{$^{56}_{28}Ni$}&{$^{100}_{50}Sn$}&{$^{132}_{50}Sn$}&{$^{208}_{82}Pb$}&\\
\hline
%$1s1/2$&$-30.73$&$ -35.08 $&$-41.18$&$-44.25$&$-37.63$&$-40.03$&$1s1/2$\\
%$1p3/2$&$-17.85$&$-26.76 $&$-33.14$&$-38.28$&$-32.72$&$-36.25$&$1p3/2$\\
%$1p1/2$&$-12.74$&$-24.78  $&$-31.32$&$-37.33$&$-32.10$&$-35.89$&$1p1/2$\\
%$1d5/2$&$-5.25$&$ -17.75 $&$-24.33$&$-31.49$&$-27.06$&$-31.78$&$1d5/2$\\
$\textcolor{blue}{2s1/2}$&$-3.03$&${-14.19}$&${-20.51}$&$-29.27$&$-25.57$&$-30.90$&$\textcolor{blue}{1d3/2}$\\
$\textcolor{blue}{1d3/2}$&${2.11}$&$-13.58$&${-20.35}$&$-28.34$&$-24.46$&$-29.67$&$\textcolor{blue}{2s1/2}$\\
$1f7/2$&$\ldots$&$-8.33$&$-14.96$&$-24.04$&$-20.78$&$-26.74$&$1f7/2$\\
$\textcolor{blue}{2p3/2}$&$\ldots$&$-4.90 $&$-10.56$&$-20.08$&$-18.07$&$-25.07$&$\textcolor{blue}{1f5/2}$\\
$\textcolor{blue}{1f5/2}$&$\ldots$&$\textcolor{red}{-1.87}$&$-8.41$&$-19.75$&$-17.15$&$-23.59$&$\textcolor{blue}{2p3/2}$\\
$2p1/2$&$\ldots$&$\textcolor{red}{-3.00} $&$-8.32$&$-18.19$&$-16.06$&$-22.87$&$2p1/2$\\
$1g9/2$&$\ldots$&$1.16 $&$-5.25$&$-16.07$&$-14.01$&$-21.20$&$1g9/2$\\
$\textcolor{blue}{2d5/2}$&$\ldots$&$\ldots$&$-1.70$&$\textcolor{red}{-10.01}$&$-9.79$&$-18.50$&$\textcolor{blue}{1g7/2}$\\
$\textcolor{blue}{3s1/2}$&$\ldots$&$\ldots$&$-0.94$&$\textcolor{red}{-11.17}$&$-9.76$&$-17.20$&$\textcolor{blue}{2d5/2}$\\
$\textcolor{blue}{1g7/2}$&$\ldots$&$\ldots$&$3.52$&$\textcolor{red}{-8.36}$&$-7.99$&$-15.76$&$\textcolor{blue}{2d3/2}$\\
$\textcolor{blue}{2d3/2}$&$\ldots$&$\ldots$&$\ldots$&$\textcolor{red}{-9.05}$&$-7.74$&$-15.44$&$\textcolor{blue}{3s1/2}$\\
$1h11/2$&$\ldots$&$\ldots$&$\ldots$&$-7.69$&$-6.83$&$-15.24$&$1h11/2$\\
$\textcolor{blue}{2f7/2}$&$\ldots$&$\ldots$&$\ldots$&$-2.95$&$\textcolor{red}{-0.94}$&$-11.28$&$\textcolor{blue}{1h9/2}$\\
$\textcolor{blue}{3p3/2}$&$\ldots$&$\ldots$&$\ldots$&$-1.53$&$\textcolor{red}{-2.61}$&$-10.64$&$\textcolor{blue}{2f7/2}$\\
$\textcolor{blue}{3p1/2}$&$\ldots$&$\ldots$&$\ldots$&$-0.53$&$\textcolor{red}{-0.57}$&$-8.90$&$\textcolor{blue}{1i13/2}$\\
$\textcolor{blue}{1h9/2}$&$\ldots $&$\ldots$&$\ldots$&$0.59$&$\textcolor{red}{-1.34}$&$-8.45$&$\textcolor{blue}{3p3/2}$\\
$\ldots$&$\ldots$&$\ldots$&$\ldots$&$\ldots$&$0.03$&$-8.36$&$2f5/2$\\
$\ldots$&$\ldots$&$\ldots$&$\ldots$&$\ldots$&$\ldots$&$-7.55$&$3p1/2$\\
$\ldots$&$\ldots$&$\ldots$&$\ldots$&$\ldots$&$\ldots$&$-4.04$&$2g9/2$\\
$\ldots$&$\ldots$&$\ldots$&$\ldots$&$\ldots$&$\ldots$&$-3.49$&$1i11/2$\\
$\ldots$&$\ldots$&$\ldots$&$\ldots$&$\ldots$&$\ldots$&$-2.24$&$1j15/2$\\
$\ldots$&$\ldots$&$\ldots$&$\ldots$&$\ldots$&$\ldots$&$-2.07$&$3d5/2$\\
$\ldots$&$\ldots$&$\ldots$&$\ldots$&$\ldots$&$\ldots$&$-1.41$&$4s1/2$\\
$\ldots$&$\ldots$&$\ldots$&$\ldots$&$\ldots$&$\ldots$&$-1.01$&$2g7/2$\\
$\ldots$&$\ldots$&$\ldots$&$\ldots$&$\ldots$&$\ldots$&$-0.82$&$3d3/2$\\
\hline
$\chi^2$&$0.24$&$1.18$&$0.09$&$0.06$&$0.03$&$0.06$&$\ldots$\\
\end{tabular}
\end{ruledtabular}
\end{table}
%%%%%%%%%%%%%%%%%%%%%%%%%%%%%%%%%%%
%%%%%%%%%%%%%%%%%%%%%%%%%%%%%%%%%%%
\begin{table}[H]
\begin{ruledtabular}
\caption{ Single particle shell model energy states for \textbf{Proton} of doubly magic nuclei $^{16}_{8}O$, $^{48}_{20}Ca$, $^{56}_{28}Ni$, $^{100}_{50}Sn$, $^{132}_{50}Sn$  and $^{208}_{82}Pb$ obtained by our calculations using Matrix Numerov method}
\label{Table 3}
\begin{tabular}{p{1.2cm}p{1.5cm}p{1.5cm}p{1.5cm}p{1.5cm}p{1.5cm}p{1.2cm}p{1.2cm}}
{States}&\multicolumn{6}{c}{Numerical energy values(MeV)}&{States}\\
&{$^{16}_{8}O$}&{$^{48}_{20}Ca$}&{$^{56}_{28}Ni$}&{$^{100}_{50}Sn$}&{$^{132}_{50}Sn$}&{$^{208}_{82}Pb$}&\\
\hline
%$1s1/2$&$-26.68$&$-37.31 $&$-30.42$&$-28.07$&$-38.33$&$-32.70$&$1s1/2$\\
%$1p3/2$&$-14.23$&$ -28.85 $&$-22.99$&$-22.85$&$-33.65$&$-29.47$&$1p3/2$\\
%$1p1/2$&$-9.14$&$-26.48 $&$-21.06$&$-21.77$&$-32.83$&$-28.95$&$1p1/2$\\
%$1d5/2$&$-2.11$&$ -19.57  $&$-14.72$&$-16.69$&$-28.04$&$-25.37$&$1d5/2$\\
$\textcolor{blue}{2s1/2}$&$-0.21$&$ -15.18 $&$-10.71$&$-14.25$&$-26.11$&$-24.16$&$\textcolor{blue}{1d3/2}$\\
$\textcolor{blue}{1d3/2}$&${4.82}$&$-14.55 $&$-10.60$&$-13.05$&$-24.61$&$-22.26$&$\textcolor{blue}{2s1/2}$\\
$1f7/2$&$\ldots$&$ -9.75 $&$-5.89$&$-9.83$&$-21.70$&$-20.56$&$1f7/2$\\
$\textcolor{blue}{2p3/2}$&$\ldots$&$-4.98$&$-1.56$&$-5.58$&$-18.24$&$-18.37$&$\textcolor{blue}{1f5/2}$\\
$\textcolor{blue}{2p1/2}$&$\ldots$&$-2.38 $&${0.54}$&$-5.14$&$-16.97$&$-16.25$&$\textcolor{blue}{2p3/2}$\\
$\textcolor{blue}{1f5/2}$&$\ldots$&$-1.71 $&$\ldots$&$-3.56$&$-15.58$&$-15.36$&$\textcolor{blue}{2p1/2}$\\
$1g9/2$&$\ldots$&$0.40$&$\ldots$&$-2.43$&$-14.77$&$-15.18$&$1g9/2$\\
$\textcolor{blue}{2d5/2}$&$\ldots$&$\ldots$&$\ldots$&$\textcolor{red}{3.92}$&$-9.37$&$-11.71$&$\textcolor{blue}{1g7/2}$\\
$\textcolor{blue}{1g7/2}$&$\ldots$&$\ldots$&$\ldots$&$\textcolor{red}{2.55}$&$-9.10$&$-9.89$&$\textcolor{blue}{2d5/2}$\\
$1h11/2$&$\ldots$&$\ldots$&$\ldots$&$\ldots$&$-7.32$&$-9.29$&$1h11/2$\\
$\textcolor{blue}{3s1/2}$&$\ldots$&$\ldots$&$\ldots$&$\ldots$&$-6.56$&$-8.09$&$\textcolor{blue}{2d3/2}$\\
$\textcolor{blue}{2d3/2}$&$\ldots$&$\ldots$&$\ldots$&$\ldots$&$-6.43$&$-7.57$&$\textcolor{blue}{3s1/2}$\\
$\textcolor{blue}{2f7/2}$&$\ldots$&$\ldots$&$\ldots$&$\ldots$&$-1.18$&$-4.26$&$\textcolor{blue}{1h9/2}$\\
$\textcolor{blue}{1h9/2}$&$\ldots$&$\ldots$&$\ldots$&$\ldots$&$0.330$&$-3.26$&$\textcolor{blue}{2f7/2}$\\
$\ldots$&$\ldots$&$\ldots$&$\ldots$&$\ldots$&$\ldots$&$-2.95$&$1i13/2$\\
$\ldots$&$\ldots$&$\ldots$&$\ldots$&$\ldots$&$\ldots$&$-0.37$&$2f5/2$\\
$\ldots$&$\ldots$&$\ldots$&$\ldots$&$\ldots$&$\ldots$&$-0.28$&$3p3/2$\\
\hline
$\chi^2$&$1.53$&$0.31$&$0.48$&$0.25$&$0.03$&$0.11$&$\ldots$\\
\end{tabular}
\end{ruledtabular}
\end{table}
%%%%%%%%%%%%%%%%%%%%%%%%%%%%%%%%%%%
%%%%%%%%%%%%%%%%%%%%%%%%%%%%%%%%%%
Next, $J^\pi$ assignments for nuclei near to doubly magic nuclei in mass range equal to and less than $^{208}_{82}Pb$ based on our simulation, which match with experimentally\cite{15} available energy sequence, are tabulated. These assignments if we consider only the energy level sequence of lighter nuclei only, are compared with those calculated using energy level sequence given in usually referred textbook \textit{Introductory Nuclear Physics} by Kenneth Krane\cite{4} and are shown in Table \ref{Table 4}, showing that we cannot consider same energy level sequence for lighter and heavier nuclei. The discrepancies in spin assignments are highlighted in blue colour. These discrepancies are also observed in other textbooks \cite {5,6,17,22,23} as well. 
%%%%%%%%%%%%%%%%%%%%%%%%%%%%%%%%%%%%%%%%%%%%%%%%%%%%%%%%%%%%%%%%%%%%%%
\begin{table}[H]
\begin{ruledtabular}
\caption{ Nuclear single particle shell model states for \textbf{Neutron} and \textbf{Proton} of nuclei near to closed shell nuclei, according to Ref.$\cite{4}$ and Current work.}
\label{Table 4}
\begin{tabular}{p{1.9cm}r p{1.9cm}r p{1.9cm}r p{1.9cm}r p{1.7cm}r p{1.7cm}r}
{Nuclei}&\multicolumn{2}{c}{Neutron states}&{Nuclei}&\multicolumn{2}{c}{Proton states}\\
&{Ref.\cite{4}}&{Current work}&&{Ref.\cite{4}}&{Current work}\\
\hline
$^{17}_{8}O$&$1d5/2$&$1d5/2$&$^{16}_{9}F$&$1d5/2$&$1d5/2$\\
$^{41}_{20}Ca$&$1f7/2$&$1f7/2$&$^{40}_{21}Sc$&$1d3/2$&$1d3/2$\\
$^{49}_{20}Ca$&$2p3/2$&$2p3/2$&$^{48}_{21}Sc$&$1f7/2$&$1f7/2$\\
$^{57}_{28}Ni$&$2p3/2$&$2p3/2$&$^{56}_{29}Cu$&$1f7/2$&$\textcolor{blue}{2p3/2}$\\
$^{101}_{50}Sn$&$1g7/2$&$\textcolor{blue}{2d5/2}$&$^{100}_{51}Sb$&$1g9/2$&$\textcolor{blue}{2d5/2}$\\
$^{133}_{50}Sn$&$1h9/2$&$\textcolor{blue}{2f7/2}$&$^{132}_{51}Sb$&$1h11/2$&$\textcolor{blue}{2d3/2}$\\
$^{209}_{82}Pb$&$2g9/2$&$\textcolor{blue}{1i11/2}$&$^{208}_{83}Bi$&$1h9/2$&$\textcolor{blue}{2f7/2}$\\
\end{tabular}
\end{ruledtabular}
\end{table}
%%%%%%%%%%%%%%%%%%%%%%%%%%%%%%%%%%%
%%%%%%%%%%%%%%%%%%%%%%%%%%%%%%%%%%
Hence, from our observations, the level sequence for lighter mass range and heavy mass range nuclei can be modified accordingly so as to provide students with data consistent with experiments.
\section{Conclusion:}
The time-independent Schr{\"o}dinger equation (TISE) for a nucleus modeled using Woods-Saxon potential along with spin-orbit coupling term has been solved numerically by choosing Matrix Numerov method. The main advantage of matrix Numerov method is that it can be easily extended to any arbitrary potential of interest. It is implemented in Gnumeric worksheet environment to obtain numerical solutions of single-particle neutron and proton states for doubly magic nuclei $^{40}_{20}Ca$. Then, using guided enquiry strategy, a Constructivist approach, students were grouped and encouraged to obtain energy level sequences for other doubly magic nuclei up-to $Z = 82$,  i.e  $^{16}_{8}O$, $^{48}_{20}Ca$, $^{56}_{28}Ni$, $^{100}_{50}Sn$, $^{132}_{50}Sn$ and $^{208}_{82}Pb$. Based on these obtained level structures, students obtained ground state total angular momentum and spin assignment for various nuclei close to doubly magic nuclei successfully and it could easily be extended to test the limits of validity.


\newpage
\section{References}
\begin{thebibliography}{30}
\bibitem{1}Eugene Meyer, “Shell model of nuclear structure,” Am. J. Phys. 36, 250–257 (1968)
\bibitem{2}R. D. Woods and D. S. Saxon, "Diffuse Surface Optical Model for Nucleon-Nuclei Scattering". Physical Review. 95 (2): 577–578(1954).
\bibitem{3}\url{https://www.ugc.ac.in/pdfnews/7870779_B.SC.PROGRAM-PHYSICS.pdf}.
\bibitem{4}Krane, S.Kenneth, \textit{Introductory Nuclear Physics}(Jon Wiley \& Sons, New York, 1988).
\bibitem{5}G.N. Ghoshal,\textit{Atomic and Nuclear Physics}(S. Chand and Co, 1997).
\bibitem{6} Samuel S.M. Wong, \textit{Introductory nuclear physics}(New Jersey: Prentice Hall, 1990).
\bibitem{7}Jithin Bhagavathi, Swapna Gora, V.V.V. Satyanarayana, O. S. K. S. Sastri and, B.P. Ajith, ``Gamma Spectra of Non-Enriched Thorium Sources using PIN Photodiode and PMT based Detectors'' Physics Education 36, 1-15(2020).
\bibitem{8} B.P. Jithin, V.V.V. Satyanarayana, S. Gora, O. S. K. S. Sastri and, B.P. Ajith, ``Measurement Model of an Alpha Spectrometer for Advanced Undergraduate Laboratories,'' Physics Education 35, 1-14(2019).
\bibitem{9}Swapna Gora, B.P. Jithin, V.V.V. Satyanarayana, O.S.K.S Sastri, and B.P Ajith, ``Alpha Spectrum of 212 Bi Source Prepared using Electrolysis of Non-Enriched ThNO3 Salt,'' Physics Education 35, 1-16(2019).
\bibitem{10} Aditi Sharma, Swapna Gora, Jithin Bhagavathi, and O. S. K. S Sastri, "Simulation study of nuclear shell model using sine basis", Am. J. Phys. 88, 576 (2020).
\bibitem{11}O.S.K.S Sastri, Aditi Sharma, Jyoti Bhardwaj, Swapna Gora, Vandana Sharda and Jithin B.P,``Numerical Solution of Square Well Potential With Matrix Method Using Worksheets,'' Physics Education, 1-14(2019).
\bibitem{12}Aditi Sharma and O. S. K. S. Sastri, "Numerical simulation of quantum anharmonic oscillator, embedded within an infinite square well potential, by matrix methods using gnumeric spreadsheet", European Journal of Physics (2020):1-20.
\bibitem{13}O. S. K. S. Sastri, Aditi Sharma, Shikha Awasthi, Anil Khachi and Lalit Kumar,``Simulation of Vibrational Spectrum of Diatomic Molecules Using Morse Potential by Matrix Methods in Gnumeric Worksheet,'' Physics Education, Accepted for publication (2020).
\bibitem{14} Mohandas Pillai, Joshua Goglio, and Thad G. Walker, "Matrix Numerov Method for Solving Schrödinger’s Equation", American Journal of Physics. 80, 1017 (2012).
\bibitem{15} N. Schwierz, I. Wiedenhover, and A. Volya, “Parameterization of the Woods-Saxon potential for shell-model calculations,” preprint arXiv:0709.3525 (2007).
\bibitem{16}David Hestenes, ``Toward a modeling theory of physics instruction,'' Am. J. Phys. 55, 440-454(1987).
\bibitem{17} Kris LG. Heyde, \textit{The nuclear shell model} ( Springer, Berlin, Heidelberg, 1994).
\bibitem{18} Aage Bohr and Ben R. Mottelson, Nuclear Structure (World Scientific,Singapore, 1998).
\bibitem{19}\url{http://www.chemistry.wustl.edu/~dgs/chem436/Lectures/Chem-436-\\Lecturenotes.pdf} for solving Woods-Saxon potential using Runge-Kutta method.
\bibitem{20} "Guided inquiry strategy" [Online accessed 9-July-2016 from \url{https://icwc.wikispaces.com/file/view/Guided+Inquiry.doc}]
\bibitem{21}\url{https://saivyasa.in/moodle/message/index.php?id=2}
\bibitem{22} Atam.P.Arya, Fundamentals of Nuclear Physics (Allyn and Bacon, Inc. Boston,1966).
\bibitem{23} Casten, R., and Richard F. Casten. Nuclear structure from a simple perspective.Vol. 23. Oxford University Press on Demand, 2000.
\end{thebibliography}
\end{document}